\documentclass[reprint,amsmath,amssymb,aps,pra, superscriptaddress]{revtex4-2}
\usepackage{graphicx}
\usepackage{dcolumn}
\usepackage{bm}
\usepackage[hidelinks]{hyperref}
\hypersetup{colorlinks=true, urlcolor=blue, linkcolor=blue, citecolor=blue}

\begin{document}

\title{A Mosquito-Inspired Theoretical Framework for Acoustic Signal Detection}

\author{Justin Faber}
\affiliation{Department of Physics \& Astronomy, University of California, Los Angeles, California, USA}
\email{faber@physics.ucla.edu}

\author{Alexandros C Alampounti}
\affiliation{University College London Ear Institute: London, UK}

\author{Marcos Georgiades}
\affiliation{University College London Ear Institute: London, UK}

\author{Joerg T Albert}
\affiliation{University College London Ear Institute: London, UK}
\affiliation{Cluster of Excellence Hearing4all, Sensory Physiology \& Behaviour Group, Department for Neuroscience, School of Medicine and Health Sciences, Carl von Ossietzky Universität Oldenburg, Oldenburg, Germany}

\author{Dolores Bozovic}
\affiliation{Department of Physics \& Astronomy, University of California, Los Angeles, California, USA}
\affiliation{California NanoSystems Institute, University of California, Los Angeles, California, USA}

\date{\today}

\begin{abstract}
Distortion products are tones produced through nonlinear effects of a system simultaneously detecting two or more frequencies. These combination tones are ubiquitous to vertebrate auditory systems and are generally regarded as byproducts of nonlinear signal amplification. It has previously been shown that several species of infectious-disease-carrying mosquitoes utilize these distortion products for detecting and locating potential mates. It has also been shown that their auditory systems contain multiple oscillatory components within the sensory structure, which respond at different frequency ranges. Using a generic theoretical model for acoustic detection, we show the signal-detection advantages that are implied by these two detection schemes: distortion product detection and cascading a signal through multiple layers of oscillator elements. Lastly, we show that the combination of these two schemes yields immense benefits for signal detection. These benefits could be essential for male mosquitoes to be able to identify and pursue a particular female within a noisy swarm environment.
\end{abstract}

\maketitle

\begin{figure*}[t]
\centering
\includegraphics[width=17.8cm]{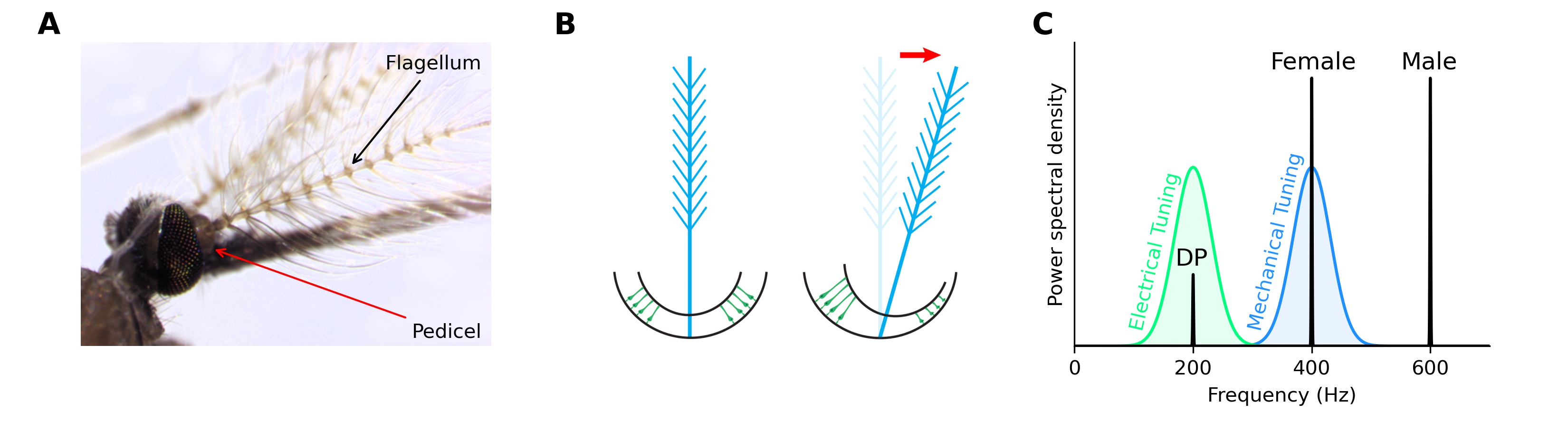} 
\caption{(\textbf{Mosquito Hearing}) (A) The feather-like flagellum and the bowl-shaped pedicel, which houses the Johnston's organ (male \textit{Anopheles gambiae}). (B) Illustration of the mechanical connection between the flagellum (blue) and the active sensory neurons (green). A deflection of the flagellum causes an extension of the sensory elements. (C) Approximate tuning ranges of the mechanical components of the organ (blue) and the electrical response of the neurons (green) for male mosquitoes. Female and male wingbeat frequencies are indicated, along with resulting distortion product (DP).}
\label{fig:intro}
\end{figure*}

Mosquitoes pose a significant public health problem, as they spread pathogens that cause more than 700,000 deaths each year \cite{WHO2020}. Furthermore, predictions based on climate change models warn that the geographic areas which are vulnerable to mosquito-borne diseases such as malaria, dengue fever, and others, will greatly spread in the coming decades \cite{messinaManyProjectedFutures2015}. Meanwhile, traditional means of controlling the mosquito population, utilizing chemicals such as DTT, have been decreasing in efficacy as resistance develops and have been proven to be detrimental to the environment \cite{WHO2018}. An understanding of the mechanisms of auditory detection in mosquitoes could enable alternate means of controlling mosquito populations, as the male mosquitos rely on their sense hearing to locate potential mates \cite{andresBuzzkillTargetingMosquito2020, suAssessingAcousticBehaviour2020}. However, many basic questions remain open in our understanding of how a mosquito performs sensitive and highly frequency-tuned detection of sound. 

Despite their largely different structures, the auditory systems of insects exhibit several parallels to those of vertebrates, which have been much more extensively studied \cite{albertComparativeAspectsHearing2016, warrenBridgingGapMammal2021}. First, both systems have been shown to involve mechano-electrical transduction, a process by which vibrations induced by the incoming sound wave cause mechanical gating of the associated ion channels, thus converting a mechanical signal to an electrical one. Second, the sense of hearing in both insects and vertebrates is characterized by highly nonlinear response functions, with power-law growth of response to stimuli of increasing amplitude \cite{martinCompressiveNonlinearityHair2001, gopfertSpecificationAuditorySensitivity2006, gopfertHearingInsects2016, nadrowskiAntennalHearingInsects2011}. Third, both systems reveal the presence of an energy-consuming amplification process. In certain species, this process can manifest itself in active limit cycles: some vertebrates exhibit innate hair-bundle oscillations \cite{Martin2001, Martin2003, martinActiveHairbundleMove2016} and spontaneous otoacoustic emissions \cite{hudspethIntegratingActiveProcess2014}, while male mosquitoes display naturally-occurring self-sustained oscillations of their flagella (Fig. \ref{fig:intro}A-B)\cite{gopfertActiveAuditoryMechanics2001, suSexSpeciesSpecific2018}. Fruit flies possess similar machinery and can exhibit chemically-induced self-sustained flagellar oscillations, though this active system naturally resides in a quiescent state \cite{gopfertPowerGainExhibited2005}.

Nonlinear dynamics models have provided an extensive theoretical framework for the study of vertebrate hearing. Specifically, theoretical models based on the normal form equation for the Hopf bifurcation have been shown to capture many of the key phenomena that characterize the auditory system \cite{choeModelAmplificationHairbundle1998, eguiluzEssentialNonlinearitiesHearing2000}. The compressive nonlinearity of the response, amplification of low-intensity sound, frequency selectivity, and the presence of active oscillation are all well described by this simple equation, showing consistency across many measurements performed on different species. 

Important differences, however, arise between the nonlinear dynamics underlying insect hearing, compared to those established to hold for vertebrates. In all of the vertebrate systems studied, the frequency characterizing innate oscillation coincides with that of optimal sensitivity of detection. In insects, however, these frequencies can differ \cite{warrenSexRecognitionMidflight2009, pennetierSingingWingMechanism2010, simoesMaskingAuditoryBehaviour2018, suSexSpeciesSpecific2018}. Male mosquitoes of several species have been shown to exhibit self-sustained oscillations (SSOs) of their flagella, with oscillation frequencies comparable to the wingbeat frequency of the corresponding females. As these SSOs have been shown to synchronize with female wingbeats, they are believed to serve as an amplification and filtering mechanism, enhancing the male's ability to detect a female and thus improving his chances of mating. Surprisingly, however, the optimal tuning of the male's sensory neurons does not coincide with the frequency of interest - female wingbeat frequency - or the male's own SSO frequency \cite{lapshinFrequencyTuningIndividual2013, lapshinFrequencyOrganizationJohnston2017}. Instead, the optimal frequency of detection of the sensory neurons was shown to coincide with a distortion product corresponding to a combination of the male's own wingbeat with that of the female (Fig. \ref{fig:intro}C) \cite{warrenSexRecognitionMidflight2009, simoesMaskingAuditoryBehaviour2018, suSexSpeciesSpecific2018}. 

Distortion products (DPs), also referred to as phantom tones, are a generic feature of nonlinear systems and are ubiquitous in all active amplifiers. They have previously been observed in vertebrates at the level of individual hair cells \cite{barralPhantomTonesSuppressive2012a} as well as \textit{in vivo}; upon presentation of two stimulus waves, these phantom tones are manifested as additional peaks in the spectra, which correspond to linear combinations of the two applied frequencies \cite{roblesTwotoneDistortionBasilar1991}. As vertebrate systems are dominated by a cubic nonlinearity, the highest peaks in the distortion spectra have been shown to be third order. However, in vertebrates, these distortions are a byproduct of the amplification mechanism, whose main purpose is to enhance the detection of primary tones (PTs). 

The auditory system of mosquitoes seems not to be poised to detect the primary tones, but rather the distortion products. Specifically, the sensory neurons exhibit a characteristic frequency that aligns with a combination tones of male and female wingbeat frequencies \cite{warrenSexRecognitionMidflight2009, simoesMaskingAuditoryBehaviour2018, suSexSpeciesSpecific2018}. While primary tone detection has been extensively studied, to the best of our knowledge, no model has yet explored the effects of distortion-product detection. In the current work, we use a generic model for this auditory system to show that detecting distortion products of the stimulus, as opposed to the primary tones (PTs), results in substantial enhancement in the frequency selectivity of the system. As frequency discrimination and selectivity are likely to be essential for reliable communication, we speculate that this mechanism is important for auditory detection by mosquitoes. We show the phenomenon to be ubiquitous, applicable across many orders of distortion tones, and constitutes a general design principle for sharpening the frequency selectivity of an active detector.

Detection poised at combination tones, however, does include a significant trade-off, in that tuning is sharpened at the expense of sensitivity of detection. As insects must be capable of detecting faint signals, this limitation would prove harmful. We hence explore the possibility that amplifiers are nested, with the output of one active oscillator providing the input to another. Thus, we treat the flagellum and the mechanically sensitive neuronal elements as two separate active elements, with the incoming sound stimulating the flagellum, and the response of the flagellum providing stimulus to the neurons. We demonstrate that cascading the signal through several layers of detectors enhances amplification for weak and near-resonance stimuli, while providing greater attenuation of large-amplitude and off-resonance signals. The sensitivity enhancement improves with additional elements in the cascade, but with diminishing returns. 

Finally, we combine the effects of DP detection and signal cascading in a configuration that mimics the male mosquito's auditory system. We consider a two-oscillator cascade, with the first oscillator representing the flagellum and the second representing the mean field of the neuronal elements. We evaluate the signal-detection capabilities of this composite system in terms of sensitivity to weak stimuli, attenuation of strong stimuli, and frequency selectivity for the female wingbeats. For all three metrics, the composite system performs better than either of the two oscillators alone, suggesting that DP detection may be utilized by male mosquitoes in order to enhance their ability to detect and pursue a female.

While the model is inspired by recent experimental observations of the mosquito auditory system \cite{warrenSexRecognitionMidflight2009, simoesMaskingAuditoryBehaviour2018, suSexSpeciesSpecific2018}, the findings are more general. We propose that cascading of active nonlinear elements can lead to significant improvement in the sensitivity of a detector. Further, if one of the elements in the cascade in tuned to a distortion product rather than primary tones, frequency selectivity can be vastly enhanced.

\section*{Results}

We first explore the full features of distortion product detection, by considering an individual active oscillator tuned to one of the combination tones of a two-tone stimulus. Second, we explore how recursively cascading the response of one oscillator into another affects signal detection to a single, near-resonance tone. Each active element in the cascade is represented by a normal form equation for the Hopf oscillator. Finally, we combine these two features to propose a model of the auditory system of the mosquito. We consider the simplest case, in which there are only two oscillators in the cascade, and the second oscillator is tuned to a cubic distortion product of the male and female wingbeat. Generalizations of this model are also discussed.

\subsection*{Distortion-product tuned detectors}

\begin{figure*}[t]
\centering
\includegraphics[width=17.8cm]{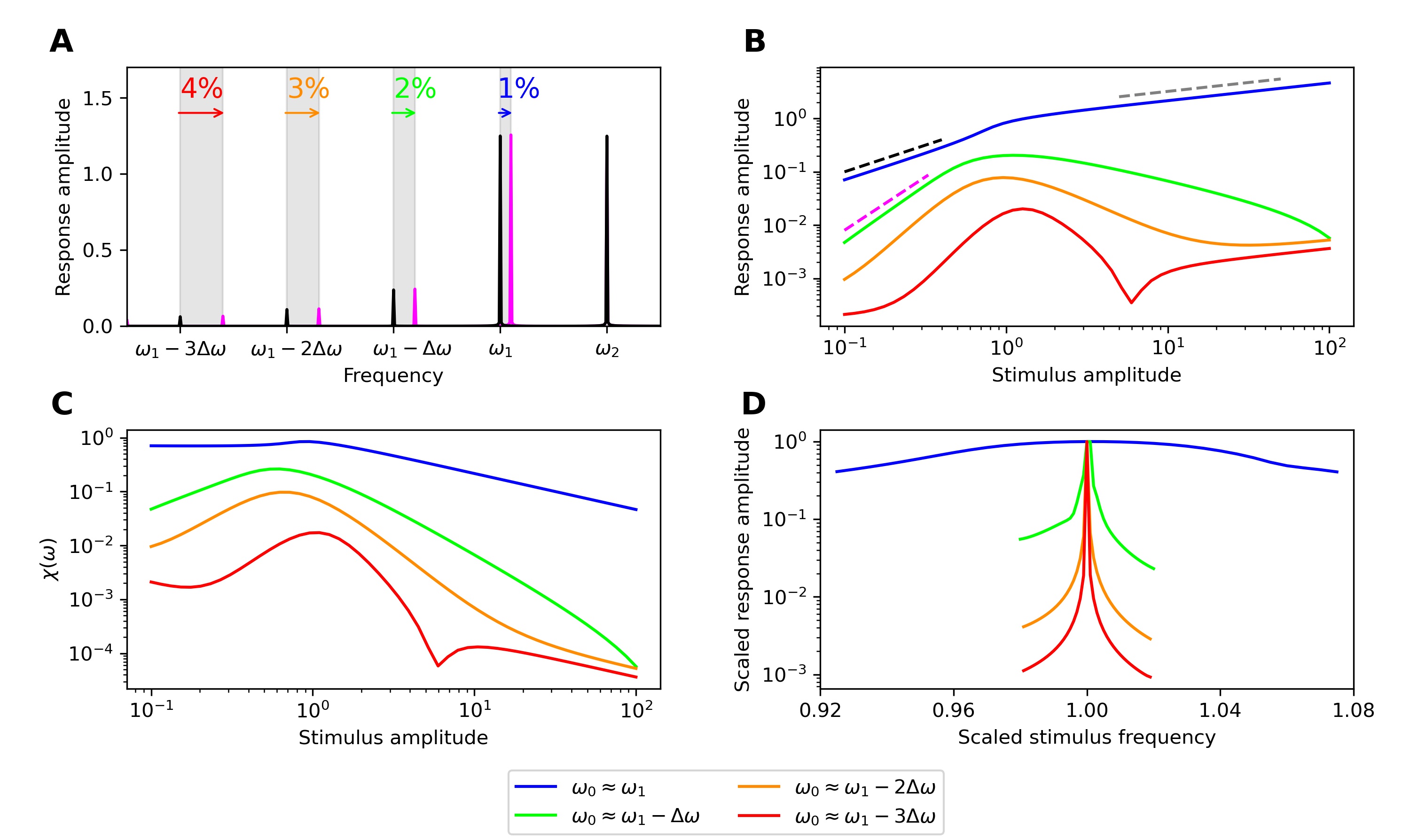} 
\caption{(\textbf{Distortion-Product Detection}) (A) Fourier transform of the response of a Hopf oscillator to a two-tone stimulus (black curve). The arrows illustrate the frequency shifts associated with increasing $\omega_1$ by 1\% (pink curve). Phase-locked response amplitude (B) and linear response function (C) of a single Hopf oscillator exposed to a range of stimulus amplitudes, $F_1$. Dashed lines represent power law growth with exponents of 1/3 (grey), 1 (black), and 2 (pink). (D) Scaled response amplitude to a frequency sweep in $\omega_1$. For panels (A) and (B), $F_2 = 5$ and $\frac{\omega_2}{\omega_1} = 1.1$. For panel (D), $\omega_1$ is modulated around $2\pi$, while $\omega_2= 1.1 \times 2\pi$, and $F_1 = F_2 = 0.1$. For all panels, $\mu=0.1$.}
\label{fig:dp}
\end{figure*}

\begin{figure*}[t]
\centering
\includegraphics[width=17.8cm]{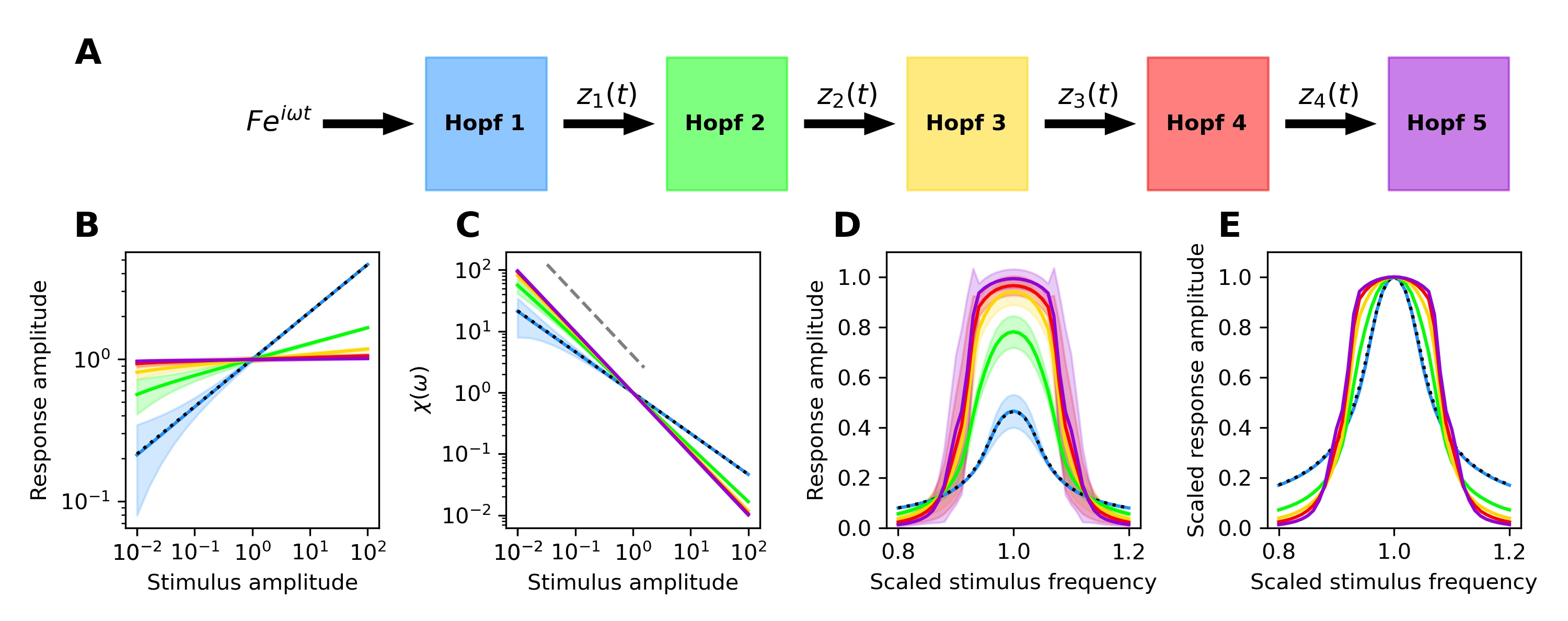}
\caption{(\textbf{Amplifier Cascade}) (A) Schematic of the external signal passing through five layers of detection. Response amplitude (B) and linear response function (C) at the stimulus frequency ($\omega=\omega_0$) for a range of stimulus amplitudes. The grey, dashed line corresponds to $\frac{1}{F}$ dependence. (D) Response amplitude for a range of stimulus frequencies with $F=0.1$. (E) Same as (D) but with curves normalized to their peaks at $\omega=\omega_0$. The colors of the curves correspond to the layer of cascaded as illustrated in (A). All $\mu_j$ values were chosen pseudo-randomly from a Gaussian distribution with mean $0$ and standard deviation $0.1$. The solid curves and shaded regions correspond to the mean and standard deviation of each measure from $10$ unique combinations of random $\mu_j$ values. For comparison, the black, dotted curves correspond to a single Hopf oscillator tuned precisely to the bifurcation point, $\mu_1 = 0$.}
\label{fig:cascade}
\end{figure*}

We first demonstrate the effects of tuning an active detector to one of the distortion product frequencies of a two-tone stimulus. We consider a system described by a complex variable, $z(t)$, which is governed by the normal form equation for the supercritical Hopf bifurcation \cite{izhikevichDynamicalSystemsNeuroscience2007},

\begin{eqnarray}
\frac{dz}{dt} = (\mu + i\omega_0)z - |z|^2z + F_1 e^{i\omega_1 t} + F_2 e^{i\omega_2 t},
\label{eq:hopf1}
\end{eqnarray}

\noindent where $F_1$ and $F_2$ are the stimulus amplitudes, $\omega_1$ and $\omega_2$ are the stimulus frequencies, and $\mu$ represents the control parameter of the system. For $\mu < 0$, the system resides in the quiescent regime, while for $\mu > 0$, the system displays self-sustained oscillations at the characteristic frequency, $\omega_0$. The Hopf bifurcation occurs at the critical point between these two regimes ($\mu=0$). Without loss of generality, we let $\omega_2 \geq \omega_1$.

This system exhibits nonlinear responses at the distortion product frequencies,

\begin{eqnarray}
\omega_{p, q} = p\omega_1 - q\omega_2,
\label{eqn:dp}
\end{eqnarray}

\noindent where $p$ and $q$ are integers, and the distortion product order is defined as the sum of the magnitudes ($|p| + |q|$). The cubic term in Eq. \ref{eq:hopf1} results in distortion products of only odd order. The distortion products lower than $\omega_1$ can be represented as $\omega_{p, q} = \omega_1 - \frac{1}{2}(|p|+|q|-1)\Delta\omega$, where $\Delta\omega = \omega_2 - \omega_1 > 0$. It has been demonstrated that distortion-product amplitudes fall off exponentially with increasing order \cite{julicherPhysicalBasisTwotone2001}. In Fig. \ref{fig:dp}A, we illustrate this effect. The figure further demonstrates that the sensitivity to modulations in the stimulus frequency, $\omega_1$, increases with increasing distortion-product order. We propose that this effect could serve to enhance frequency selectivity and frequency discrimination of external tones. This can be understood, in part, through calculating the changes in distortion-product frequency with respect to $\omega_1$,

\begin{eqnarray}
\frac{d\omega_{p, q}}{d\omega_1} = p,
\label{eqn:dp2}
\end{eqnarray}

\noindent which increases with increasing DP order. Note that we are considering $\omega_1$ to be the stimulus tone of interest, while examining the low-frequency distortion products. However, these effects can also be seen for modulation in $\omega_2$ and for high-frequency distortion products.

We chose two stimulus tones that differ minimally in frequency, $\frac{\omega_2}{\omega_1} = 1.1$, so as to observe several orders of distortion products. We calculate the phase-locked response amplitude and the linear response function, $\chi(\omega)$ (see Methods), over a range of stimulus amplitudes, $F_1$ (Fig. \ref{fig:dp}B-C). Naturally, the response amplitudes and sensitivity decrease for increasing orders of the distortion product. The oscillator tuned to the primary tone exhibits the expected linear growth for small stimulus amplitudes and compressive response with $1/3$ power law for large stimulus amplitudes \cite{dukeCriticalOscillatorsActive2008}. 

However, when the detector is tuned to a distortion product, the response grows at least quadratically for weak stimulus inputs and becomes extremely compressive for large stimulus amplitudes, with responses even decreasing with increasing $F_1$ (see Appendix for an analytic approximation of this response). This enhanced growth at weak stimulus levels could enable an insect to resolve variations in the volume of faint sounds, while the extreme compression at higher amplitudes could maintain a large dynamic range. This alternate detection scheme hence exhibits advantages over the traditional response function of an oscillator tuned to the primary tone, but it entails a reduction of the overall sensitivity. Further, the non-monotonic nature of the response curves implies an ambiguity in identifying stimulus levels, as there is not a one-to-one relationship of input and output amplitudes. This ambiguity at the level of an individual oscillator could, however, be resolved in an array of such detectors.

The most striking benefit of tuning a detector to a distortion product emerges when calculating the frequency selectivity of the response. Notably, the quality factor of the response increases by orders of magnitude when the system is tuned to a distortion product (see Appendix for an analytic calculation). The frequency selectivity continues to increase with increasing distortion-product orders (Fig. \ref{fig:dp}D). This feature arises as a result of several effects. First, modulations in the primary tones are magnified at the distortion product level (Eq. \ref{eqn:dp2}). This magnification, in turn, leads to a more narrow range of frequencies over which the detector responds. Second, large stimulus amplitudes that would normally saturate the response to primary tones are attenuated at the distortion product level, due to the previously described compression. These effects result in sharper tuning curves and a larger quality factor.

\subsection*{Cascade of nonlinear amplifiers}

Next, we establish the effects of recursively cascading the response of one Hopf oscillator as input into another. To explore these effects independently of those of distortion-product detection, we apply a single-tone stimulus to the system and measure the phase-locked response to the signal, applied at or near resonance. For simplicity, we assume each Hopf oscillator in the series to have the same characteristic frequency, $\omega_0$. The dynamics of the first oscillator are described by

\begin{eqnarray}
\frac{dz_1}{dt} = (\mu_1 + i\omega_0)z_1 - |z_1|^2z_1 + F e^{i\omega t}
\label{eqn:cascade1}
\end{eqnarray}

\noindent and the dynamics of the following elements in the series are described by

\begin{eqnarray}
\frac{dz_j}{dt} = (\mu_j + i\omega_0)z_j - |z_j|^2z_j +  z_{j-1}(t),
\label{eqn:neuron}
\end{eqnarray}

\noindent where $\omega_0=2\pi$, and the indexing starts at $j=2$ and terminates at $j=j_{max}$. To test for robustness of the cascading scheme, we allow the control parameters to deviate from the bifurcation point. Prior to simulating the dynamics, each $\mu_j$ is chosen pseudo-randomly from a Gaussian distribution centered at $0$ with standard deviation $0.1$.

Cascading  a signal through multiple elements enhances several of the features exhibited by individual Hopf oscillators. At low amplitudes and in the vicinity of the characteristic frequency, the nested oscillators exhibit higher amplification of the signal, increasing the response and the sensitivity. Further, the system exhibits stronger compression of large-amplitude stimuli. When a single Hopf oscillator receives a large-amplitude stimulus, the response grows as $F^{\frac{1}{3}}$, since the response is dominated by the cubic term. For a cascading system receiving a strong stimulus, the response of the $j^{th}$ element grows as $F^{(\frac{1}{3})^j}$. After several layers of cascading, the response flattens out and approaches a constant amplitude (Fig. \ref{fig:cascade}B), and the sensitivity approaches a $\frac{1}{F}$ dependence (Fig. \ref{fig:cascade}C).

Further, cascading of Hopf oscillators was observed to amplify only near-resonance stimuli, while attenuating off-resonance frequencies. Thus, the narrow range of frequencies in the immediate vicinity of $\omega_0$ is enhanced, with a sharp decrease and vanishing tail ends as frequency is varied (Fig. \ref{fig:cascade}D-E). The frequency selectivity of the system is therefore enhanced in comparison to that of an individual element.

We note that the benefits of the proposed cascade of Hopf oscillators exhibit different features than what is observed by simply tuning one oscillator more precisely to the bifurcation point (dotted curves in Fig. \ref{fig:cascade}). In Fig. \ref{fig:cascade}B-E, we compare the sensitivity and frequency selectivity of the response. As can be seen, the enhancement of sensitivity, compression, and frequency selectivity of the response is improved as more layers are added to the cascade (increasing $j_{max}$). Most of the enhancement, however, already occurs in the first several layers, with diminishing returns upon continuing the cascade. Further, the benefits are robust to imperfect tuning of the control parameters.

\subsection*{Generic model of the mosquito's auditory system}

\begin{figure*}[t]
\centering
\includegraphics[width=17.8cm]{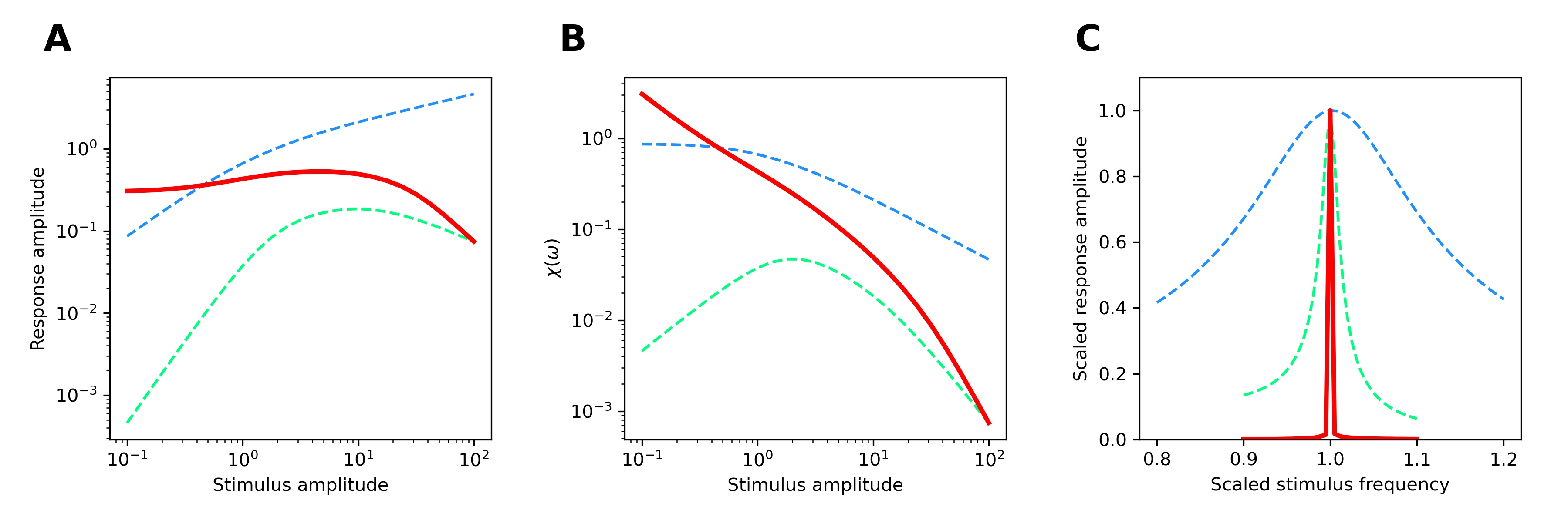}
\caption{(\textbf{Mosquito Model}) Phase-locked response amplitude (A) and linear response function (B) to a two-tone stimulus ($\omega_f = \omega_0$). (C) Scaled amplitude response of the system to a range of female stimulus frequencies ($F_f = 0.1$). Solid red curves represent the responses to the composite, two-oscillator detector, while blue and green dashed curves represent responses of single-oscillator systems comprised of only the first and only the second oscillators, respectively. For all panels, $F_m =5$, $\omega_0 = 2\times2\pi$, $\omega_m = 3\times2\pi$,  and $\mu_1 = \mu_2 = 0.1$.}
\label{fig:composite}
\end{figure*}

Finally, we combine the two effects described previously in order to build a simple model for signal detection by male mosquitoes. The feather-like flagellum at the end of the antenna of a mosquito oscillates in response to an incoming sound wave. The base of the antenna contains the Johnston's organ, which has a bowl-like structure with rotational symmetry. The pivots of the antenna modulate the open probability of the transduction channels embedded in the ciliated sensory neurons that connect to the antennal base (Fig. \ref{fig:intro}A-B). To construct a simplified model of these elegant structures, we describe the flagellar position by a time-dependent complex variable, $z_1(t)$. For simplicity, we describe the mean field of the active neuronal elements by a second complex variable, $z_2(t)$. Both of these state variables are governed by the normal form equation for the supercritical Hopf bifurcation,

\begin{eqnarray}
\frac{dz_1}{dt} = (\mu_1 + i\omega_0)z_1 - |z_1|^2z_1 + F_f e^{i\omega_f t} + F_m e^{i\omega_m t}
\label{eqn:cascade1}
\end{eqnarray}

\begin{eqnarray}
\frac{dz_2}{dt} = [\mu_2 + i(2\omega_0 - \omega_m)]z_2 - |z_2|^2z_2 +  z_1(t),
\label{eqn:neuron}
\end{eqnarray}

\noindent where the sinusoidal forcing terms represent the input stimuli, with $F_f$ and $F_m$ denoting the amplitudes and $\omega_f$ and $\omega_m$ the frequencies of the female and male wingbeats, respectively. The male flagellum exhibits a characteristic frequency $\omega_0$, while $\mu_1$ represents the control parameter of the flagellum. For $\mu_1 < 0$, the flagellum is in the quiescent state, while for $\mu_1 > 0$, the flagellum exhibits self-sustained oscillations powered by an internal energy source. This control parameter, $\mu_1$, thus determines the dynamic state of the flagellum.

Although several orders of distortion products may be important for signal detection, we consider only the simplest case, where the neuronal elements are tuned to the cubic distortion product between the characteristic frequency of the flagellum and the male's wingbeat frequency, $2\omega_0 - \omega_m$. We use experimental data obtained from prior studies on several mosquito species to approximate the relationship between the frequencies \cite{somersHittingRightNote2022}. Hence, we consider a stimulus frequency ratio of $\frac{\omega_m}{\omega_f} \approx 1.5$, and set the male flagellum to be tuned near the female wingbeat frequency $\omega_0 \approx \omega_f$, which makes the neuronal elements tuned near the cubic distortion product frequency, $2\omega_0 - \omega_m \approx 2\omega_f - \omega_m$ (Eq. \ref{eqn:neuron}), consistent with the experimental results.

We compare the signal detection properties of this two-oscillator system to each of the two oscillators alone. This composite system exhibits sensitivity to weak stimuli that is greater than either individual oscillator (Fig. \ref{fig:composite}A-B). Further, the composite system displays amplitude compression at high stimulus amplitudes that is comparable to the second, more compressive, oscillator. Most strikingly, the composite system is drastically more frequency selective than either of the other two individual oscillators (Fig. \ref{fig:composite}C).

\section*{Discussion}

Prior numerical models of mosquito hearing have accounted for several of the experimentally measured phenomena of the mosquito's auditory system, including self-sustained oscillations of the flagellum, the nonlinear and hysteretic response function that results from stimulus amplitude ramps, and the appearance of a frequency-doubling component at the neuronal level. These phenomena were described using a microscopic integrate-and-twitch model \cite{avitabileMathematicalModellingActive2009} that tracks the dynamics of the individual neuronal elements, which in turn apply impulsive forcing back to the flagellum. This model has been further extended to the continuum limit \cite{champneysMECHANICSHEARINGCOMPARATIVE2011}. A phenomenological model has also been developed, in which the flagellum is assumed to be a simple harmonic oscillator, while the active neuronal units are represented either by a single Hopf oscillator representing the mean field response \cite{jacksonSynchronyTwicefrequencyForcing2009}, or an array of coupled Hopf oscillators, each representing a cluster of active elements \cite{windmillFrequencyDoublingActive2018}. 

With the current study, we construct a general model for mosquito hearing, based on dynamical systems theory, aimed to capture the basic features of this remarkable auditory system. While the normal form equation for the Hopf bifurcation, tuned for optimal detection of a primary tone, has been shown to describe all the key characteristics of the vertebrate system, it does not readily generalize to that of insects. We aim here to develop a framework, which can then be fine-tuned to describe specific behaviors of different mosquito species, and possibly those of other flagellar insects. We propose that two key features emerge in constructing a general model: cascading of nonlinear amplifying elements and detection of distortion products. 

Using a generic model of active auditory detection, we have assessed the advantages and disadvantages of distortion-product tuning as opposed to the more intuitive primary-tone tuning. DPs contain less energy than the primary tones from which they are generated. As a result, DP detection comes at the cost of reduced overall sensitivity. However, certain remarkable advantages arise. First, the response function of the DP detector exhibits more curvature than that of the PT detector, showing more rapid growth at low-level stimuli and more intense compression for strong stimuli. Together, these effects yield a higher resolution to the intensities of faint tones, while minimizing the sensitivity to loud tones.

The most noteworthy difference between the two forms of detection lies in the frequency selectivity of the response. DP detectors display much sharper frequency tuning, with quality factors increasing with increasing DP order. The degree to which the quality factor is improved for a cubic DP detector increases and diverges with weaker stimulus amplitudes (see Appendix). This enhancement is a consequence of two mechanisms. First, the DP detector attenuates any incoming, off-resonance signals, narrowing the range in frequencies over which detection will occur. Second, DP frequencies are sensitive to modulations in the PT frequency, with sensitivity growing with increasing DP order. This results in a further narrowing of this detection range. We also demonstrated that this sharpening of the response is not specific to the cubic term, but arises in all orders of distortion product, increasing at higher orders. A single Hopf oscillator acting as a PT detector can also exhibit strong frequency selectivity when poised at the bifurcation. The quality factor of the response diverges as the forcing amplitude goes to zero. However the DP detector produces a larger quality factor for non-zero forcing, and which diverges more rapidly than that of the PT detector for vanishing stimulus amplitudes (see Appendix).

Inspired by the auditory system of the mosquito, we explored the effects of connecting several Hopf oscillators in a cascade, with the response of each oscillator acting as the stimulus for the next. We note that this detection scheme differs from arrays of active Hopf oscillators arranged tonotopically, which have previously been used to model the mammalian cochlea \cite{magnascoWaveTravelingHopf2003}. The configuration used in our study is an attempt to mimic the multiple components of the mosquito's auditory system. Sound stimulates the flagellum, which then stimulates the neurons. There may be additional levels of cascading at the neuronal level, which could be readily accounted for by additional elements in the cascade. We found that cascading enhances sensitivity to weak stimuli, while more significantly compressing the response to large-amplitude signals. Further, each level of detection distorts the shape of the response curve, making the system more sensitive to near-resonance stimulus and less sensitive to off-resonance stimulus, while negligibly affecting the quality factor.

Lastly, we constructed a system that exhibits both of the described effects by feeding the response of a PT detector into a DP detector. We chose characteristic frequencies to reflect those observed in experimental measurements of mosquitoes. This generic 2-oscillator signal detector exhibits high sensitivity to weak stimuli, significant compression of strong stimuli, and extreme frequency selectivity. Further, all three characteristics are stronger for the 2-oscillator system than for either of the two components alone.

We conceptualize the mosquito's auditory system as a mechanical resonator feeding its response into an electrical oscillator of different characteristic frequency. Intuitively, this mismatch in frequency tuning should attenuate inputs of any frequency and yield poor performance as a signal detector. However, using a generic model for auditory detection and considering the nonlinear distortion products of the response, we motivate this counter-intuitive detection scheme, and explain the empirical observations in experimental studies \cite{warrenSexRecognitionMidflight2009, simoesMaskingAuditoryBehaviour2018, suSexSpeciesSpecific2018}. We show that this class of systems responds sensitively to weak stimulus, while attenuating strong signals. Most strikingly, this scheme yields immense frequency selectivity, which may be essential for the male mosquito to isolate the flight tone of a female within a noisy swarm environment. Future experimental work will have to test the predicted signal-analytical gains of our theoretical explorations more directly. Here, especially the roles and molecular nature of the elusive control parameter, $\mu$, will be of interest. It is tempting to speculate that it is linked to the reported vast efferent neuromodulation of the mosquito ear \cite{andresAuditoryEfferentSystem2016, georgiadesHearingMalariaMosquitoes2023}. Intriguingly, both distortion products and efferent innervation are poorly understood features of human hearing, too.

\section*{Numerical Methods}

Numerical simulations were performed using the fourth order Runge-Kutta method with time steps of $10^{-3}$ and approximately $10^5$ time steps to ensure high frequency resolution.

\subsection*{Response amplitude and linear response function}

Phase-locked amplitudes were calculated by computing the Fourier transform of the response,

\begin{eqnarray}
\tilde{z}(\omega) = \int^\infty_{-\infty} z(t)e^{-i\omega t} dt,
\label{eqn:ft}
\end{eqnarray}

\noindent and taking the magnitude of the Fourier component either at the stimulus frequency for primary-tone detectors ($|\tilde{z}(\omega_f)|$), or at the distortion product of interest for distortion-product-tuned detectors ($|\tilde{z}(\omega_{p, q})|$).

To estimate the sensitivity to a particular stimulus, we calculate the magnitude of the linear response function,

\begin{eqnarray}
\chi(\omega) = \bigg|\frac{\tilde{z}(\omega)}{\tilde{F}(\omega_f)}\bigg| = \frac{|\tilde{z}(\omega)|}{F_f},
\label{eqn:ft}
\end{eqnarray}

\noindent at the frequency of interest, where $\tilde{F}(\omega)$ is the stimulus in frequency space.

\newpage

\section*{Acknowledgments}

This work was supported by a grant from the Biotechnology and Biological Sciences Research Council, UK (BBSRC, BB/V007866/1 to J.T.A.) and a grant from The Human Frontier Science Program (HFSP grant RGP0033/2021 to J.T.A. and D.B.). The authors thank Judit Bagi and Anwen Bullen for providing high-quality mosquito photographs.

\bibliography{Bibliography}

\section*{Appendix}

\subsection*{Quality factor of the response to single-tone stimulus}

Here we estimate the quality factor of the response of a Hopf oscillator to a single-tone stimulus. We use

\begin{eqnarray}
\frac{dz}{dt} = (\mu + i\omega_0)z - |z|^2z + F(t),
\label{eqn:Hopf}
\end{eqnarray}

\noindent where $F(t) = Fe^{i\omega t + \psi}$ is the stimulus with phase offset, $\psi$. We assume a response at the stimulus frequency of the form

\begin{eqnarray}
z(t) = Re^{i(\omega t + \phi)}.
\end{eqnarray}

\noindent Inserting this ansatz into Eq. \ref{eqn:Hopf} and multiplying each side by its complex conjugate results in a cubic equation for the response amplitude,

\begin{eqnarray}
F^2 = (R^3 - \mu R)^2 + (\omega - \omega_0)^2 R^2.
\label{eqn:1tone_response}
\end{eqnarray}

\noindent The response amplitude grows as $R \approx F^{\frac{1}{3}}$ for stimuli close to the characteristic frequency and for strong stimuli. However, for stimuli far from the characteristic frequency, the response can be approximated as $R \approx \frac{F}{|\omega-\omega_0|}$ if $\mu$ is small. We approximate the response near resonance up to second order in ($\omega-\omega_0$), which gives the response amplitude as a function of the frequency detuning,

\begin{eqnarray}
R \approx F^{\frac{1}{3}} - \frac{(\omega - \omega_0)^2}{6F},
\label{eqn:1tone_approx}
\end{eqnarray}

\noindent for systems close to the bifurcation ($|\mu| \ll 1$). This parabola can be used to estimate the width of the tuning curve at half its maximum in energy. The characteristic frequency divided by this width defines the quality factor of the response. For a single-tone stimulus, the approximation of the quality factor yields

\begin{eqnarray}
Q_{\text{PT}} = \frac{1}{\sqrt{24(1-\frac{1}{\sqrt{2}})}} \bigg( \frac{\omega_0}{F^{2/3}} \bigg).
\label{eqn:Q1}
\end{eqnarray}

\subsection*{Quality factor of distortion-product response}

We now approximate the quality factor of a Hopf oscillator tuned near the cubic distortion-product frequency ($\omega_0 \approx \omega_{dp} = 2\omega_f - \omega_m$). The forcing term takes the form $F(t) = F_fe^{i(\omega_f t + \psi_f)} + F_me^{i(\omega_m t + \psi_m)}$. We use the frequency ratio described in the main text $\frac{\omega_m}{\omega_f} \approx 1.5$ and assume a response of the form

\begin{eqnarray}
\begin{split}
z(t) = R_fe^{i(\omega_f t + \phi_f)} &+ R_me^{i(\omega_m t + \phi_m)} \\ &+ R_{dp}e^{i(\omega_{dp} t + \phi_{dp})},
\end{split}
\end{eqnarray}

\noindent where we are considering only the three largest frequency contributions. Inserting this into Eq. \ref{eqn:Hopf} and separating terms according to their frequency contribution gives rise to three equations for the three amplitudes. 

Responses at the primary tones can be easily estimated from the equations corresponding to their respective frequency components. Since both tones are far from the resonance frequency, the responses can be estimated as

\begin{eqnarray}
R_f \approx \frac{F_f}{\Delta\omega} \text{    and    } R_m \approx \frac{F_m}{2\Delta\omega},
\end{eqnarray}

\noindent provided that the forcing is sufficiently weak. Isolating contributions to the distortion-product frequency gives

\begin{eqnarray}
\begin{split}
i\omega_{dp}R_{dp} = (\mu + i\omega_0)R_{dp} &- R_{dp}(2R_f^2 + 2R_m^2 +R_{dp}^2) \\ &+ R_f^2R_me^{(2\phi_f-\phi_m-\phi_{dp})}.
\end{split}
\end{eqnarray}

\noindent Computing the magnitude of this relationship yields

\begin{eqnarray}
\begin{split}
R_f^4R_m^2 = (\delta\omega R_{dp})^2 + (R_{dp}^3 &+ 2R_f^2R_{dp} \\ &+ 2R_m^2R_{dp} - \mu R_{dp})^2,
\end{split}
\end{eqnarray}

\noindent where $\delta\omega = \omega_{dp} - \omega_0 = 2\omega_{f} - \omega_{m} - \omega_0$. Note that modulations in $\omega_{f}$ are magnified by a factor of $2$ in the distortion product. This equation for $R_{dp}$ can be written in the same form as the single-tone case,

\begin{eqnarray}
F_{\text{eff}}^2 = (R_{dp}^3 - \mu_{\text{eff}} R_{dp})^2 + (\delta\omega)^2 R_{dp}^2,
\label{eqn:1tone_response}
\end{eqnarray}

\noindent where $F_{\text{eff}} = R_f^2R_m$ and $\mu_{\text{eff}} = \mu - 2(R_f^2 + R_m^2)$ are the effective forcing and effective control parameter, respectively \cite{stoopTwoToneSuppressionCombination2004}. Notice that the effective forcing grows as $F_f^2$. This explains the quadratic growth in the distortion-product response seen in Fig. \ref{fig:dp}B.

Since this equation for $R_{dp}$ has the same form as the single-tone relationship, we can use the same second-order expansion as described for the single-tone case if this system resides near the effective bifurcation point ($|\mu_{\text{eff}}| \ll 1$). We compute the quality factor with respect to variations in the distortion-product frequency,

\begin{eqnarray}
Q_{\text{DP}} = \frac{1}{\sqrt{24(1-\frac{1}{\sqrt{2}})}} \bigg( \frac{\omega_0(\Delta\omega)^2}{(\frac{1}{2}F_f^2F_m)^{2/3}} \bigg).
\label{eqn:Q2}
\end{eqnarray}

When comparing $Q_{DP}$ to $Q_{PT}$, a factor of $2$ is gained because modulations in $\delta\omega$ correspond to twice those in the female stimulus frequency. However, this factor of $2$ cancels with the factor of $2$ loss from the distortion-product detector being tuned to half the frequency of the primary tone. Taking the female tone to be the frequency of interest in the single-tone case, we can estimate by how much the quality factor is improved from detecting the first cubic distortion product instead of the primary tone,

\begin{eqnarray}
\frac{Q_{\text{DP}}}{Q_{\text{PT}}} =  \frac{(\omega_m - \omega_f)^2}{(\frac{1}{2}F_f F_m)^{2/3}},
\label{eqn:Qratio}
\end{eqnarray} 

\noindent which continues to grow for weaker stimulus levels.

\end{document}